\documentclass[twocolumn]{aastex631}

\usepackage{amsmath} 

\usepackage{hyperref}
\usepackage{verbatim}
\usepackage[normalem]{ulem}

\newcommand\req{R_{\rm eq}}
\newcommand\rpol{R_{\rm pol}}

\newcommand\rref{r_{\rm ref}}
\newcommand\rrref{R_{gr}}

\shorttitle{Giant Planet Shapes}
\shortauthors{Mankovich et al.}

\begin{document}

\title{Can radio occultations constrain Uranus or Neptune's internal rotation periods?}

\author{Christopher R. Mankovich}
\author{Alex B. Akins}
\author{Dustin Buccino}
\affiliation{Jet Propulsion Laboratory, California Institute of Technology}
\author{Ravit Helled}
\affiliation{University of Z\"urich}
\author{Marzia Parisi}
\affiliation{Jet Propulsion Laboratory, California Institute of Technology}

\begin{abstract}
The shapes of fluid planets bear the signatures of rotational flattening and atmospheric flows.
Precise knowledge of their shapes and wind profiles may therefore reveal their interior rotation rates.
We re-examine this idea for the ice giants, where missions like the Uranus Orbiter and Probe could use radio occultations to measure {atmospheric heights near 1~bar} at multiple latitudes, complementing Voyager~2's near-equatorial constraint for Uranus.
Applying geodetic calculations and considering zonal wind uncertainties, we find that only a narrow range for Uranus's 1-bar polar radius, $\rpol=24968.6\pm4.7$~km, is consistent with Uranus's winds, occultations, and gravity field, even treating Uranus's interior spin as a free parameter.
This is because the isobaric shape depends on the total rotation of the isobaric surface, which is already well constrained by observations, irrespective of what portion is attributed to bulk rotation versus winds.
{Occultations will, however, be valuable for testing our underlying assumption that the winds manifest the full differential rotation that sets the shape. T}he apparent north-south asymmetry in Uranus's winds{, if permanent, produces} a $5$-km difference between the northern and southern polar radii, measurable with suitable radio occultations.
Neptune's much more uncertain winds yield $\sim100$~km variations in polar and equatorial radii.
We confirm that Uranus and Neptune's magnetic rotation periods yield nonzero mean dynamical heights for their {atmospheres. A}ccurate results for Uranus and Neptune require that the full latitude-dependent rotation be incorporated when fitting radii from occultations.
Only significantly faster interior rotation---periods close to 15 h in both Uranus and Neptune---would minimize their dynamical heights.

\end{abstract}

\section{Introduction}\label{sec.intro}

Fluid planets are made oblate by their rotation. This oblateness can be constrained via precise measurement of the atmospheric pressure/altitude relationship. Radio occultations proffer atmospheric refractivity data that can reveal the radii of constant-pressure surfaces within the atmosphere, typically at pressures of order 100 mbar to 1 bar (e.g., \citealt{1981JGR....86.8721L}). The planetary shapes derived from this technique are ubiquitous in the literature, constituting key inputs for models of atmospheres and interiors, as well as the coordinate systems that serve as a foundation for a variety of analyses \citep{2018CeMDA.130...22A}. New data along these lines have recently been obtained for Jupiter by Juno \citep{2025GeoRL..5213231C, 2025GeoRL..5216804S} and similar measurements would lie at the intersection of interior and atmospheric science attainable by a future Uranus Orbiter and Probe mission (UOP; \citealt{simon2021uopstudy,2024arXiv241201872H,2024SSRv..220...10D}).

Meanwhile, the internal rotation states of the solar system giants are a topic of active research. Jupiter's internal rotation has long been well-constrained from ground-based observations of Jovian radio emissions that are periodically modulated as the magnetic field rotates with the planet. 
Similar radio emissions from Uranus observed by Voyager 2 revealed a $17.239\pm0.009$~h period at Uranus, corroborated by a less precise measurement from magnetometry \citep{1986Sci...233...85N}. 
More recently, \cite{2025NatAs...9..658L} used ultraviolet auroral emissions to track Uranus's magnetic poles over more than a decade, constraining an apparently highly stable magnetic rotation period to an extraordinary precision of 36 ms.
At Neptune, periodic radio emissions recorded by Voyager 2 revealed a $16.11\pm0.05$~h magnetic rotation period \citep{1989Sci...246.1498W}.

However, magnetometry from Voyager 2's encounters with Uranus and Neptune also painted a picture of magnetic fields with extraordinary complexity. Each planet features a magnetic dipole axis that is highly inclined and offset from its spin axis \citep{1986Sci...233...85N,1989Sci...246.1473N} and an unexpectedly large magnetic quadrupole moment \citep{1987JGR....9215329C,1991JGR....9619023C}, pointing toward a dynamo configuration unlike that of Jupiter, Saturn, or Earth. Theories appeal to relatively thin shell dynamos that enclose an electrically insulating or stably stratified inner region (\citealt{2006Icar..184..556S,2024PNAS..12103981M}; see \citealt{2020RSPTA.37890479S} for a review of this developing topic). With little settled about magnetic field generation or the electrical conductivity profile within Uranus or Neptune, it may be premature to rule out the possibility that magnetospheric dynamics introduce some degree of temporal variability in these planets' radio emissions, or that relatively thin and shallow dynamo regions may rotate differentially with respect to the deeper interior. Independent constraints on the planetary rotation are therefore desirable.

No clear-cut magnetospheric rotation period exists for Saturn because its rotation and magnetic dipole axes are nearly perfectly aligned \citep{2020Icar..34413541C}. Periods in Saturn's kilometric radiation vary with time (see \citealt{2008JGRA..113.5222K} and references therein). However, estimates from atmospheric dynamics \citep{2009Natur.460..608R,2007Sci...317.1384A,2009P&SS...57.1467H} and ring seismology \citep{2019ApJ...871....1M,2023PSJ.....4...59M} yield rotation periods that agree within a range of 2 to 3 minutes. Other estimates for Saturn based on fitting rigidly rotating models to Saturn's observed oblateness \citep{2015Natur.520..202H,2019ApJ...879...78M} fall in the same range, despite neglecting Saturn's strong zonal winds, which generally alter the direction of local gravity and hence the overall shape of constant-pressure surfaces in a planet's atmosphere. This agreement reflects the fact that Saturn, like Jupiter, appears to be close to a configuration in which the interior rotation and zonal wind pattern yield small dynamical heights \citep{2009P&SS...57.1467H}. 
{Section~\ref{sec.shapes} gives our formal definition of the atmospheric dynamical height; here it suffices to say that their small dynamical heights simply mean that} Jupiter and Saturn's physical shapes are each well approximated by a rigidly rotating constant-potential surface rotating at the planet's intrinsic rotation period. 
{There is no general, physical mechanism that mandates this behavior; see relevant discussion in \cite{2007Sci...317.1384A}, \cite{2009Natur.460..608R}, \cite{2010Icar..210..446H}, and Section~\ref{sec.discussion} below.}
{In any case, the small dynamical heights at Jupiter are}
the reason why \cite{1981JGR....86.8721L}, despite neglecting Jupiter's zonal winds, arrived at a geoid shape {similar to later,}
more accurate models \citep{2009P&SS...57.1467H,2020JGRE..12506354B,2023GeoRL..5002321G,2026NatAs.tmp...29G}{, overestimating Jupiter's polar radius by just $\sim10$~km (see \citealt{2026NatAs.tmp...29G} and our Appendix~\ref{app.jupiter_and_saturn}).}

If{, whatever the reason,} Jupiter and Saturn's internal rotation periods tend to minimize the dynamical height of their atmospheres, does the same condition hold for the more distant giants, Uranus and Neptune? If it does, can this condition be used to test the premise that the magnetic rotation periods observed for Uranus and Neptune are indeed rooted in the rotation of their interiors \citep{2010Icar..210..446H}? More generally, to what degree are measurements of the planetary shape constraining for the interior rotation?

In this work we use standard geodetic calculations to confirm \cite{2010Icar..210..446H}'s finding that the magnetic rotation periods cited above for Uranus and Neptune yield substantial (60-80~km) dynamical heights at 1 bar in these planets when their zonal wind patterns are accounted for. We argue, however, that the physical isobaric shape is a function only of the total atmospheric rotation, independent of any heuristic construction that separates this rotation into a rigid-body component and a wind-induced component responsible for the dynamical heights. 

We first summarize the geodetic model\footnote{The code behind this paper's analysis and figures is maintained at \url{https://github.com/chkvch/giant-planet-geoids}. The release corresponding to this paper is archived at \url{https://doi.org/10.5281/zenodo.18023581} \citep{mankovich_2025_18023581}.} used to calculate planetary shapes (Section~\ref{sec.geodetic}), the adopted zonal wind profiles from feature tracking observations (Section~\ref{sec.wind_profiles}), and baseline results for the shapes of Uranus and Neptune (Section~\ref{sec.basic_results}). Section~\ref{sec.fit_rigid_geoid_to_data} pauses to illustrate the consequences of neglecting winds and fitting rigidly rotating reference geoids directly to occultation points.
Section~\ref{sec.minimize_dynamical_heights} addresses the question of which reference rotation periods minimize the dynamical heights of Uranus' and Neptune's atmospheres. 
Section~\ref{sec.uncertainties} identifies the leading sources of uncertainty in each planet's equatorial and polar radius, shedding light on what aspects of these planets' atmospheres might be meaningfully constrained by new occultations or new analysis of those from Voyager 2.
We discuss these results in Section~\ref{sec.discussion} before concluding in Section~\ref{sec.conclusion}.

\section{Shape model}\label{sec.shapes}
\subsection{Geodetic calculation}\label{sec.geodetic}
The effect of zonal winds is to modify the local centrifugal support and hence the effective gravity vector. This alters the shape of constant-pressure (isobaric) surfaces, which are everywhere perpendicular to the local vertical as mandated by hydrostatic balance. 
The geodetic calculation presented by \cite{1985AJ.....90.1136L} separated the shape into two components, first a rigidly rotating equipotential \textit{reference geoid}, and second the \textit{dynamical height} describing the wind-induced deviation between the reference geoid and the physical isobaric shape. This picture relies on a somewhat artificial distinction between the rigid rotation describing the reference geoid and the spatially varying part attributable to the winds. Results from observations such as feature tracking are framed in terms of wind velocity or drift rates with respect to a presumed rigid rotation rate (e.g., the system III rotation), ultimately constraining the \textit{total} rotation frequency or period as a function of latitude. Hence, we opt here to follow the geodetic calculation used by \cite{1992AJ....103..967L} and \cite{2023GeoRL..5002321G}, which directly includes the total rotation frequency $\Omega=\Omega(\phi)$ in the radial and polar components of the gravity:
\begin{align}
    g_r(r,\phi) &= - \frac{GM}{r^2}  + \frac23\Omega^2r\cos^2\phi \nonumber \\
    &+ \frac{GM}{r^2} \sum_{n=1}^\infty(2n+1)J_{2n}\left(\frac{\rrref}{r}\right)^{2n}P_{2n}(\mu), \label{eq.gravity_radial}
\end{align}
\begin{align}
g_\phi(r,\phi) = &- \frac13r\Omega^2\frac{d}{d\phi}P_2(\mu)  \nonumber\\
    & -\frac{GM}{r^2}\sum_{n=1}^\infty J_{2n}\left(\frac{\rrref}{r}\right)^{2n}\frac{d}{d\phi}P_{2n}(\mu)
     \label{eq.gravity_polar}
\end{align}
where $GM$ is the planet's gravitational mass, $J_{2n}$ are its zonal gravity coefficients, $R_{gr}$ is the reference radius used to normalize the $J_{2n}$, $\phi$ is planetocentric latitude measured from the equator with $\mu=\sin\phi$. 
{As is commonly done for the gas giants, we presume zonally symmetric rotation such that $\mathbf\Omega=\Omega\,\mathbf{\hat z}$ and the azimuthal component of the gravity is zero.
Our choice $\Omega=\Omega(\phi)$ assumes that rotation is constant on radii between the isobaric surface being considered and the surface probed by the feature-tracking observations. This is slightly different from the assumption of barotropy made by, e.g., \cite{2023GeoRL..5002321G,2025GeoRL..5216804S,2026NatAs.tmp...29G}, which would preserve $\Omega=\Omega(r\sin\phi)$ when mapping the observed rotation from an observer's reference shape onto our model's isobaric surface at the same pressure. 
Comparing the two approaches for the most oblate outer planet, Saturn, we find that the barotropic assumption modifies the best-fitting polar radius (see Appendix~\ref{app.jupiter_and_saturn}) by $\lesssim100$~m, the equatorial radius by $\sim200$~m, and the root-mean-square error to occultation points by $\sim200$~m.
}

The shape of the isobar is governed by
\begin{equation}\label{eq.shape_ode}
    \frac{dr}{d\phi}=r\tan\psi=r\frac{g_\phi}{g_r}
\end{equation}
where $\psi+\phi$ gives the planetographic latitude as a function of the planetocentric latitude $\phi$. Integration of Equation~\ref{eq.shape_ode} from a chosen polar radius $\rpol$ yields the shape $r(\phi)$ of the isobaric surface. It is this isobaric shape that may be compared directly radii estimated from atmospheric refractivity measurements made during radio occultations. 

Before we proceed, it is worth lingering on the fact that the above system is a function not of the bulk rotation or wind velocity in particular, but rather the total rotation frequency. Furthermore, because feature-tracking observations (e.g. \citealt{2015Icar..258..192S, 2018Icar..311..317T})  measure drift rates (or velocities) with regard to an assumed fixed background rotation (generally the System III rotation), it must be stressed that these observations also  ultimately constrain the \textit{total} rotation frequency. Therefore, to the degree that the gravity field $\{GM,\ J_n\}$ and atmospheric rotation $\Omega(\phi)$ are sufficiently characterized, the isobaric shape is known and can be calculated from Equations~\ref{eq.gravity_radial}-\ref{eq.shape_ode} {up to an integration constant (the polar radius, $\rpol$) that can be determined from even a single occultation at any latitude}. Put another way, the 1-bar shape is not dependent on what portion of the rotation is attributed to rigid body rotation versus winds. It can still prove instructive to discuss the dynamical heights implied by various scenarios for the reference geoid rotation, as we do below.

Gravity moments and accompanying reference radii are drawn from \cite{2024Icar..41115957F} for Uranus and \cite{2009AJ....137.4322J} for Neptune. The zonal moments with $n\geq6$ are unknown for these planets; we set them to zero by default. Increasing the leading order moment $J_6$ from zero to its end-member value ($+10^{-6}$ based on Uranus interior models \citep{2025mankovich_ugrav}) results in minor $\rpol$ variations of order 10 m.

Nowhere does the pressure enter the geodetic model explicitly. Rather, pressure enters only implicitly via the polar radius $\rpol$. In practice, radio occultations constrain the radii of the giant planets in the vicinity of 100 mbar to several bar. Varying the value of $\rpol$ to produce a satisfactory fit between $r(\phi)$ and the measured shape yields the best guess for the physical (isobaric) geoid corresponding to that pressure level, provided that the rotation profile $\Omega(\phi)$ and zonal gravity field $J_{2n}$ are known. 

\subsection{Alternative formulation}\label{sec.split_method}
For comparison's sake, we also implement the formulation of \cite{1985AJ.....90.1136L} in which the isobaric shape $r(\phi)$ is split into contributions from a rigidly rotating reference geoid $\rref(\phi)$ and the dynamical height $h(\phi)$. The latter is expressed as 
\begin{align}\label{eq.dynamical_height}
    h(\phi) = \langle g\rangle^{-1}\int_\phi^{\pi/2} 
        &u(\phi)
        \left[2\Omega_0+\frac{u(\phi)}{\rref(\phi)\cos\phi}\right] \\
        &\times \frac{\sin(\phi+\psi_{\rm ref})}{\cos\psi_{\rm ref}}\rref(\phi)\,d\phi\nonumber
\end{align}
where $\Omega_0$ is the rotation frequency of the frame in which the wind speeds $u(\phi)$ are measured, and the factor $\langle g\rangle$ is the mean magnitude of the gravity vector over the segment of field line connecting the isobaric surface to the reference isopotential surface. The reference geoid $\rref(\phi)$ can be obtained by substituting $\Omega=\Omega_0$ into Equations~\ref{eq.gravity_radial} and \ref{eq.gravity_polar} and integrating Equation~\ref{eq.shape_ode}, or by making an initial guess and relaxing to a surface of constant pseudopotential 
\begin{align}\label{eq.potential}
    U(r,\phi) = \frac{GM}{r}\Bigg[-1 &+ \sum_{n=1}^\infty J_{2n}\left(\frac{\rrref}{r}\right)^{2n}P_{2n}(\sin\phi)\Bigg] \nonumber \\
        &-\frac12\Omega_0^2r^2\cos^2\phi
\end{align}
(see \citealt{1985AJ.....90.1136L}). For the Uranus geoids fit to occultations in what follows, the two approaches yield polar radii that are consistent to within $\sim400$~m. In what follows, we use the simpler approach of Equations~\ref{eq.gravity_radial}-\ref{eq.shape_ode}. We calculate the dynamical height not from Equation~\ref{eq.dynamical_height}, but as the difference between the full isobaric shape obtained from Equations~\ref{eq.gravity_radial}-\ref{eq.shape_ode} and the shape of the reference geoid obtained by substituting $\Omega=\Omega_0$ into the same system.

\begin{figure}
    \begin{center}
        \includegraphics[width=\columnwidth]{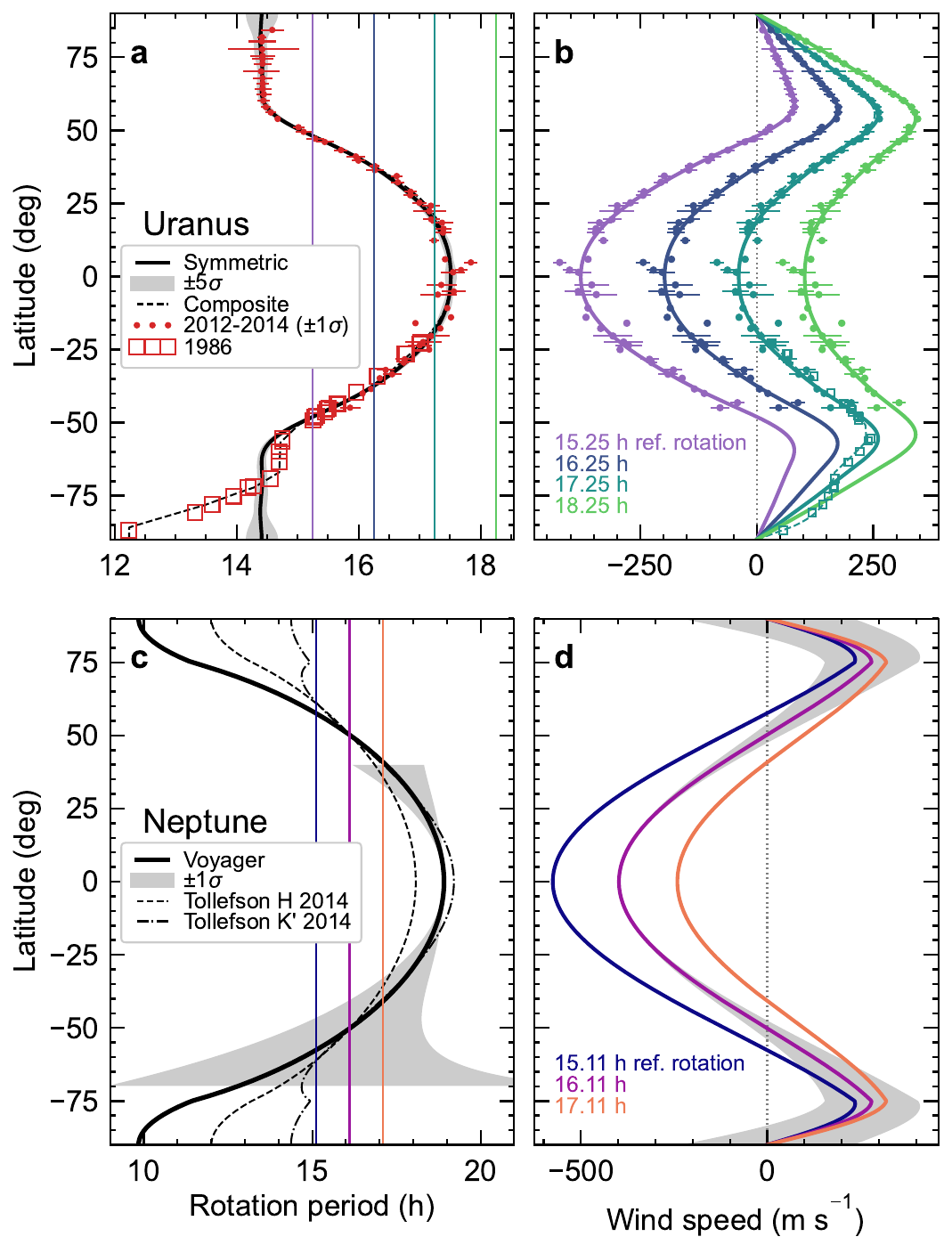}
        \caption{\label{fig.wind_profiles}
        Rotation of Uranus and Neptune's atmospheres as constrained by Voyager and ground-based imaging data.
        The profiles of atmospheric rotation period for Uranus (panel~\textbf{a}) are fits from \cite{2015Icar..258..192S}. Filled points with errors are binned data from their Tables 3-4.  The symmetric (solid) and composite (dashed) profiles differ in their north-south asymmetry, the latter incorporating Voyager imaging of the southern hemisphere (open squares) to yield a pronounced asymmetry at high latitudes. 
        The grey shaded region, most evident at the poles, shows $\pm5\sigma$ uncertainty on the symmetric fit (see text). Vertical lines show the four reference rigid rotation periods we consider for Uranus; these result in the wind speed profiles plotted in panel~\textbf{b} using the symmetric fit. Here the sole dashed profile shows the composite fit with respect to 17.25~h background rotation.
        The Neptune profiles (panel~\textbf{c}) are from \cite{2018Icar..311..317T}. Here curves show baseline fits and the grey shaded region shows the $\pm1\sigma$ spread for the Voyager fit, plotted in the latitude range over which there are data. The fits to Keck H- and K$^\prime$-band data yield larger random spreads that are not displayed here.
        Vertical lines show the three rigid rotation periods we consider for Neptune; these result in the wind speed profiles plotted in panel~\textbf{d} using the Voyager fit; the 1-sigma range for the 16.11-h reference period is plotted in grey.
        }
    \end{center}
\end{figure}

Prograde zonal flow tends to increase the height of the atmosphere; westward flow decreases it. Hence, Uranus and Neptune each exhibit competing effects between their prograde-rotating mid and upper latitudes and their retrograde equatorial regions (see Figure~\ref{fig.wind_profiles}, to be described shortly). If the magnetic rotation rates for each planet are taken to define their reference geoids, the prograde mid-latitude flows prevail in both planets, generating maximum heights of $\sim+90$~km at Uranus's 1-bar surface and $\sim+60$~km at Neptune's \citep{2010Icar..210..446H}. The relationship between the interior rotation and the magnitude of the dynamical heights is a topic of great interest and is investigated in detail in Section~\ref{sec.minimize_dynamical_heights} below.

\subsection{Zonal wind profiles}\label{sec.wind_profiles}
Figure~\ref{fig.wind_profiles} summarizes the total rotation profile and zonal wind profiles for Uranus (\citealt{2015Icar..258..192S}, panels a-b) and Neptune (\citealt{2018Icar..311..317T}, panels c-d). {For this plot alone, the sense of Uranus's rotation is reversed to adhere to the literature convention of plotting prograde rotation with positive frequency and speed. Following the IAU convention that Uranus's north pole is the one pointing (weakly) toward ecliptic north, the planetary rotation is westward.} For each planet, we show multiple profiles, as well as the statistical uncertainty associated with one of the possible profiles. These bands are obtained by randomly drawing profiles from the Gaussian-distributed polynomial fit coefficients given by Sromovsky et al.'s Table 5 (their Fit 1) and Tollefson et al.'s Table 1. Neptune's uncertainties from feature tracking are much larger than Uranus's: note that the figure displays a $\pm1\sigma$ band for Neptune (panels c-d) versus the $\pm5\sigma$ band shown for Uranus (panel a). For Uranus, \cite{2015Icar..258..192S} also provide two more fits capturing the north-south asymmetry evident at mid-latitudes in Keck/Gemini data (Asym. 5 and Asym. 6 in \citealt{2025mankovich_ugrav}). These profiles are omitted here, yielding smaller differences for our purposes than switching between the symmetric and composite fits does. For Neptune, we omit \cite{2018Icar..311..317T}'s fits to 2013 Keck data sets, which give results intermediate to the Voyager and 2014 data sets shown here.

The sources from which we take rotation profiles give longitudinal drift rates (or wind speeds) with respect to prescribed rotation periods, namely, the System III periods 17.24 h for Uranus \citep{1986Natur.322...42D} and 16.11 h for Neptune \citep{1989Sci...246.1498W}. To maintain consistency with the observed profiles, we simply calculate the total rotation frequency
\begin{equation}
    \Omega(\phi) = \omega_{\rm src}(\phi) + \Omega_{\rm src}
\end{equation}
at the outset, where $\omega_{\rm src}(\phi)$ are the drift rates from \cite{2015Icar..258..192S} (Uranus) or \cite{2018Icar..311..317T} (Neptune) and $\Omega_{\rm src}$ is the reference rigid-body rotation frequency adopted by each source. 
\cite{2018Icar..311..317T} tabulate fits for the azimuthal wind speed rather than the drift rate; these are converted to drift rates by dividing by $r_{\rm src}(\phi)\cos\phi$ where $r_{\rm src}(\phi)$ is that work's adopted reference ellipsoidal shape model. The published fits are functions of planetographic latitude; for consistency their corresponding planetocentric latitudes must be computed using the same reference shape as the source rather than our (generally different) final shape models. For our shape solutions the zonal wind speeds are then given by 
\begin{equation}
    u(\phi)=r(\phi)\cos\phi\left[\Omega(\phi) - \Omega_0\right].
\end{equation}
For the purposes of Figure~\ref{fig.wind_profiles} only, rather than making reference to our models, the shapes are approximated as ellipsoids with equatorial (polar) radii 25,559 (24,973) km for Uranus and 24,766 (24,342) km for Neptune.

At the north pole, there is no distinction between the reference geoid and the isobaric surface; the two coincide exactly because the centrifugal terms in Equations~\ref{eq.gravity_radial} and \ref{eq.gravity_polar} vanish. Likewise, in the split formulation, the final term in Equation~\ref{eq.potential} and the integral representing the dynamical height in Equation~\ref{eq.dynamical_height} both vanish. At the south pole, a small nonzero $h(-\pi/2)$ remains if the adopted wind profile is north-south asymmetric \citep{1985AJ.....90.1136L}. The strong shear seen Uranus's high southern latitudes in Voyager~2 imaging \citep{2015Icar..250..294K} stands in stark contrast with the nearly zero-shear northern latitudes recently observed near 1 bar from the ground \citep{2015Icar..258..192S}. It is an open question whether this represents a permanent asymmetry or is a manifestation of seasonal forcing{, which may lead the winds to evolve over the course of a Uranian year. The latter possibility} would suggest we may witness changes in the atmospheric dynamics as Uranus progresses through northern summer \citep{2024Icar..42016186S}. VLA observations support the existence of a transition between 20-50 bar to a deep wind that is super-rotating at $>80^\circ$ latitude with respect to the flow at $\sim1$ bar \citep{2023GeoRL..5002872A}, indicating that the deeper atmospheric flow in the north may already resemble the Voyager southern profile.

\begin{figure*}[ht]
    \begin{center}
        \includegraphics[width=0.48\textwidth]{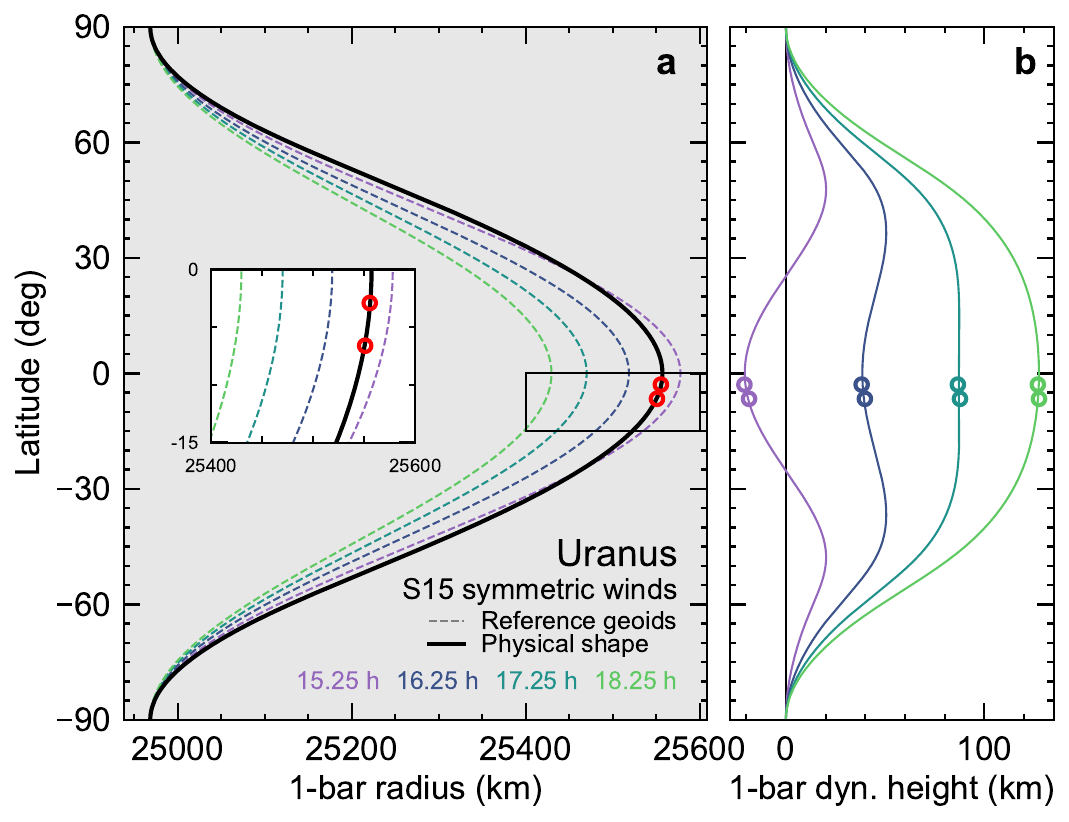}
        \hspace{0.02\textwidth}
        \includegraphics[width=0.48\textwidth]{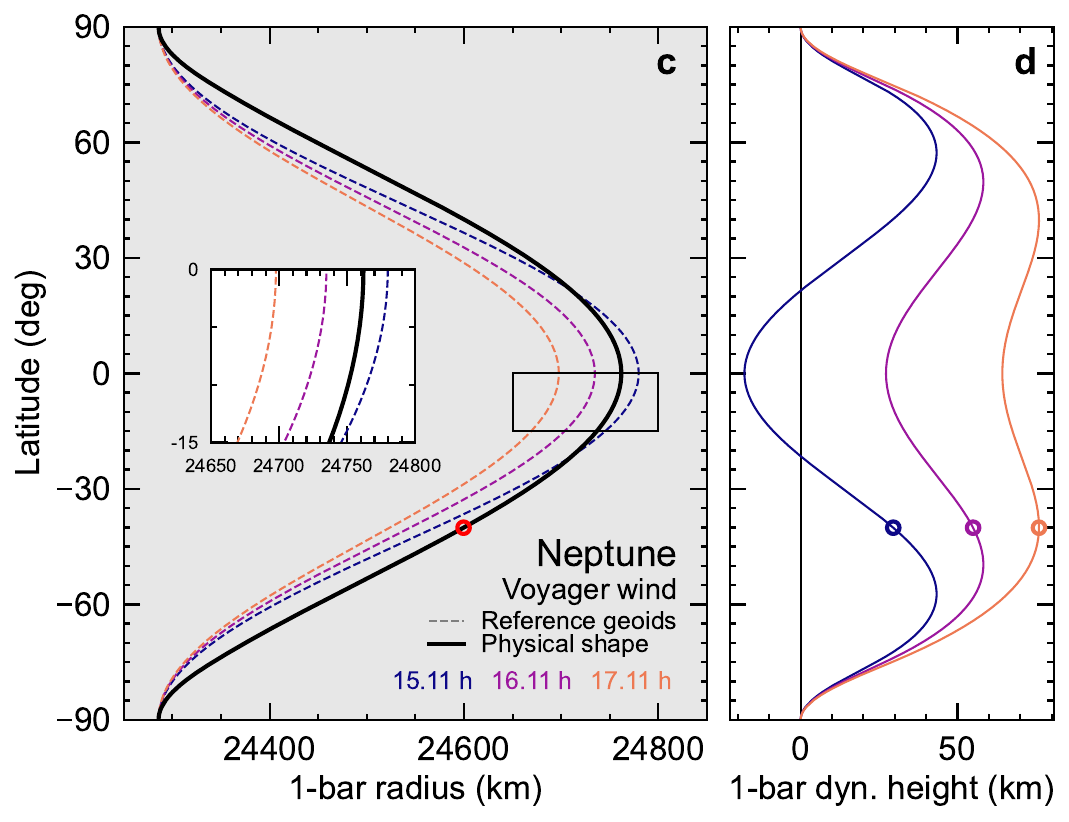}
        \caption{\label{fig.un_shapes}
        Results of fitting Uranus and Neptune's 1-bar shape (solid black curves) to 1-bar radii measured from Voyager 2 radio occultations (open circles; \citealt{1987JGR....9214987L}, \citealt{1992AJ....103..967L}),
        accounting for latitude-dependent wind heights and considering a range of underlying planetary rotation periods.
        Panel (\textbf{a}) compares the 1-bar reference geoids $\rref(\phi)$ (isopotentials assuming rigid rotation; dashed lines) to the 1-bar isobaric surface $r(\phi)$ (including modifications that zonal winds make to the local vertical; solid lines) for Uranus. 
        The inset shows detail near the equator.
        Panel (\textbf{b}) gives their difference $h(\phi)$, the dynamical height associated with the winds.
        Colors map to the rotation period $2\pi/\Omega$ assumed for the reference geoid and wind model in Equations~\ref{eq.potential}-\ref{eq.dynamical_height}.
        Dot markers give the occultation radii minus each reference geoid's 1-bar radius interpolated to the latitude of the occultation.
        Panels (\textbf{c}-\textbf{d}) give the analogous results for Neptune.
        }
    \end{center}
\end{figure*}

\subsection{Uranus and Neptune's shapes constrained by radio occultations}\label{sec.basic_results}
Figure~\ref{fig.un_shapes} shows the shapes of the isobaric surfaces (solid lines) that the above procedure produces for Uranus (panels a-b) and Neptune (panels c-d). To address the question of dynamical heights, we also show rigidly rotating reference geoids (dashed lines) for reference rotation periods within $\pm1$ to 2 hours of each planet's magnetic rotation period, taken to be 17.25 h for Uranus \citep{1986Natur.322...42D,2025NatAs...9..658L} and 16.11 h for Neptune \citep{1989Sci...246.1498W}. The dynamical heights are given by the difference between the solid curve and each of the dashed curves, and are displayed in panels (b, d).
{The geometry of the Voyager 2 Uranus occultation meant that ingress and egress were only $\sim4\deg$ apart in latitude, and hence 
the two carry limited independent information in the context of our model. We nonetheless fit both, solving for the $\rpol$ that minimizes their RMS radius error using Brent's method.}
To our knowledge the Voyager occultation radii are not explicitly given in any published source and we take them directly from Figure~8 in \cite{1987JGR....9214987L} and Figure~7 in \cite{1992AJ....103..967L}, which have sufficient resolution to locate the radii to a precision of approximately 0.1~km. 
For Uranus's zonal wind profile, we adopt \cite{2015Icar..258..192S}'s north-south symmetric fit to Keck/Gemini data. For Neptune, we adopt the nominal fit that \cite{2018Icar..311..317T} give to Voyager imaging data \citep{1993Icar..105..110S}. (In the following section we generalize to a broader family of profiles for both planets, depicted in Figure~\ref{fig.wind_profiles}). 
The Uranus profile goes to zero wind speed at the poles. For Neptune, Tollefson et al.'s even polynomial fits diverge at the poles and we follow \cite{1998Icar..136...27F} and \cite{2010Icar..210..446H} in applying a sinusoidal attenuation to zero poleward of $\pm75^\circ$ latitude to ensure physical wind speeds in these regions. 

Faster rotation generally yields more pronounced flattening.
The shape of the reference geoids is indeed found to be very sensitive to the planetary bulk rotation, as expected from Equation~\ref{eq.potential}. When Uranus's isobaric shape is fit to the near-equatorial Voyager 2 occultations, the Uranus reference geoids in Figure~\ref{fig.un_shapes}a indeed span more than a hundred km in equatorial radius, increasing from 25,430 to 25,580~km as the rotation period is decreased from 18.25 h to 15.25 h. However, this has no immediate relevance to the physical shape of the 1-bar surface, which is fixed by the observed atmospheric rotation profile $\Omega(\phi)$.
The model of Figure~\ref{fig.un_shapes}a predicts that Uranus's polar radius is $24,968.7$~km. Uranus's 1-bar equatorial radius is found to be $25,556.6$~km, nearly independent of the wind profile and gravity field because the occultation took place so close to the equator.

Our derived shape for Uranus is consistent with \cite{1987JGR....9214987L}, who advised caution in characterizing their estimate $\rpol=24,973\pm20$ km as an extrapolation from the equator; it appears that their result is robust in light of improved knowledge of Uranus's winds based on the many observations since Voyager 2. For Neptune, the Voyager wind profile implies that the 1-bar surface has an equatorial radius $24,760.6$~km and polar radius $24,285.3$~km. For the $\pm1$-h range of rigid rotation periods, the rigidly rotating reference geoids span some 80~km in equatorial radius, implying dynamical heights up to 80~km at mid-latitudes and ranging from -20 to 65 km at the equator. The 1-bar equatorial radius is in close agreement with the $24,766\pm15$~km derived by \cite{1992AJ....103..967L}. The polar radius we derive is smaller than their $24,342$~km by 57~km or 1.9 times their quoted uncertainty of $30$~km. We show below that this level of disagreement is unsurprising given the relatively large space of shape solutions generated by Neptune's considerably more uncertain wind profile. In general, our imperfect knowledge of $\Omega(\phi)$, $J_n$, and the occultation radii themselves render the polar and equatorial radii somewhat uncertain in both planets. These effects are quantified in Section~\ref{sec.uncertainties}. 

\begin{figure}
    \begin{center}
        \includegraphics[width=\columnwidth]{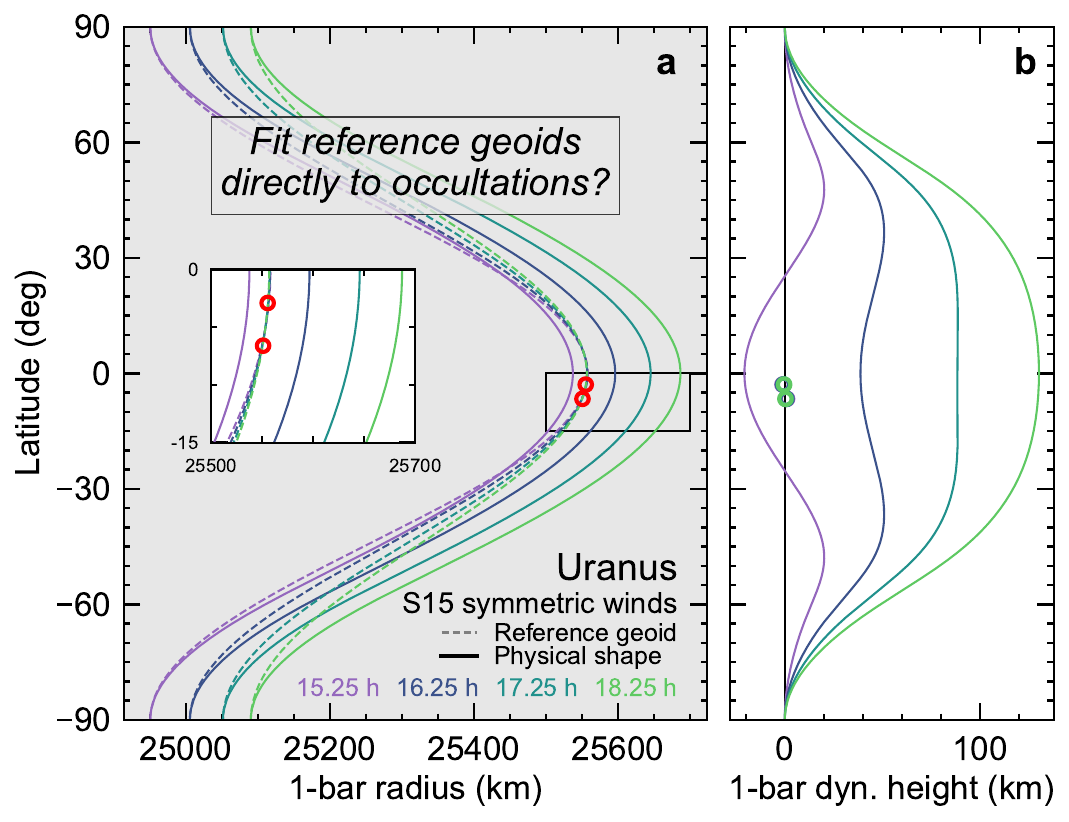}
        \caption{\label{fig.u_shape_ignore_winds}
        As in Figure~\ref{fig.un_shapes}, but here the Uranus 1-bar \textit{reference geoid} is fit to the Voyager 2 occultation radii (panel~\textbf{a}), neglecting the influence of Uranus's zonal winds. Given the nonzero dynamical heights raised by Uranus's zonal winds (see Figure~\ref{fig.un_shapes}a-b), this procedure yields 1-bar isobaric surfaces (solid curves) that fail to match occultation data and exhibit a spuriously large range of polar radii, $\sim150$~km for the periods considered here.
        }
    \end{center}
\end{figure}

\subsection{Fitting reference geoids to data: a note of caution}\label{sec.fit_rigid_geoid_to_data}
Figure~\ref{fig.u_shape_ignore_winds} takes Uranus as an example to show the result of an alternative procedure: neglecting the observed atmospheric rotation profile and instead fitting rigidly rotating reference geoids directly to occultation measurements. Here the physical radii of the 1-bar surface (solid curves) vary widely, untethered as they are to the occultation points, yielding a $\sim$150-km range for $\req$ and $\rpol$ both. At this range, Uranus's physical volume varies by nearly 2\% depending on the rotation period chosen. Each of these models fails to reproduce the occultation radii near the equator once the dynamical height of the 1-bar surface is taken into account. They therefore overrepresent the range in polar radii attainable by varying Uranus's interior rotation period. The 17.25-h model has a polar radius $\rpol=25,050.5$~km, in agreement with \cite{2010Icar..210..446H}'s $25,052\pm20$~km using the same approach. As seen in either panel of Figure~\ref{fig.u_shape_ignore_winds}, incorporating the dynamical height reveals an equatorial bulge that would overestimate Uranus's equatorial radius by more than 80~km with regard to the radio occultations (filled points).



The poor fit of the physical shape to data in Figure~\ref{fig.u_shape_ignore_winds} reflects an inconsistency that often arises when rigidly rotating interior models are adjusted so that their equatorial radii match the published values for an \emph{isobaric} geoid fit to radio occultations. It is the latter that \cite{1985AJ.....90.1136L,1987JGR....9214987L,1992AJ....103..967L} give for Saturn, Uranus, and Neptune respectively. (The exception is \citealt{1981JGR....86.8721L}, who neglected the winds for Jupiter, but reported fairly accurate results nonetheless due to Jupiter's small dynamical heights; see \citealt{2020JGRE..12506354B,2023GeoRL..5002321G}). It is these same isobaric radii that are used by the IAU to define cartographic coordinate systems (e.g., \citealt{2018CeMDA.130...22A}). However, whereas the isobaric surfaces incorporate the zonal winds, it is the rigidly rotating \textit{reference} geoids that are directly comparable to an isopotential surface from rigidly rotating interior models built on potential theory, such as concentric Maclaurin spheroids \citep{2013ApJ...768...43H} or the theory of figures \citep{1978ppi..book.....Z,2017A&A...606A.139N}. Applying the occultation-based isobaric surface's equatorial radius as an outer boundary condition for a rigidly rotating interior model introduces an error equal to the dynamical height at the equator. This discrepancy---of order 100 km or less for the solar system giants---may introduce some small, likely not critical, systematic error into elemental abundances, composition boundaries, etc. usually inferred from interior models. It is more important to understand that the resulting isopotential shapes are not directly comparable to the planet's physical shape. Any applications sensitive to the planetary shape, such as planning for future spacecraft occultations, must incorporate self-consistent modeling of the winds as described by, e.g., \cite{1985AJ.....90.1136L,2010Icar..210..446H,2023GeoRL..5002321G} and recapitulated in Section~\ref{sec.shapes}.

\section{Does the interior rotation minimize the dynamical heights?}
\label{sec.minimize_dynamical_heights}
Viewing the total velocity as effectively fixed by feature tracking observations, faster interior rotation implies more subdued prograde winds and more pronounced retrograde ones. This means that with $\rrref$, $J_{2n}$, and the total rotation profile $\Omega(\phi)$
fixed, there exists a bulk rotation rate at which the latitude-averaged dynamical height
\begin{equation}\label{eq.mean_dynamical_height}
    \langle h\rangle=\frac1\pi\int_{-\pi/2}^{\pi/2} h\,d\phi
\end{equation}
vanishes. The condition $\langle h\rangle=0$, or variations like minimizing the global maximum of $|h|$, leads to consistency with the long-known rotation period of Jupiter when applied to that planet \citep{2009P&SS...57.1467H}. Applied to Saturn, this method yields periods in the range 10h 32m to 10h 33m \citep{2007Sci...317.1384A,2015Natur.520..202H}, only slightly faster than the 10h 34m period that \cite{2009Natur.460..608R} estimated by applying a condition of marginal stability to potential vorticity profiles derived from Cassini data, and the lower end of the 10h 34m$-$10h 36m range estimated from ring seismology \citep{2023PSJ.....4...59M}. 

Although no fundamental physical argument dictates that the atmospheric flows need to obey dynamical height minimization $\langle h\rangle=0$, the general agreement that the method yields for Jupiter and Saturn raises the prospect that it might reveal the interior rotation of Uranus and Neptune, independent from the known rotation of their magnetic fields \citep{1986Natur.322...42D,1989Sci...246.1498W,2025NatAs...9..658L}. It is already clear from Figures~\ref{fig.un_shapes}b and \ref{fig.un_shapes}d that the condition $\langle h\rangle=0$ will only be met at the low range (fastest rotation) of the periods we have considered so far. A straightforward numerical optimization as a function of $\Omega_0$ shows that the mean dynamical height is minimized with $2\pi/\Omega_0=15.14$~h for Uranus and 14.48~h for Neptune. 

\begin{figure}
    \begin{center}
        \includegraphics[width=\columnwidth]{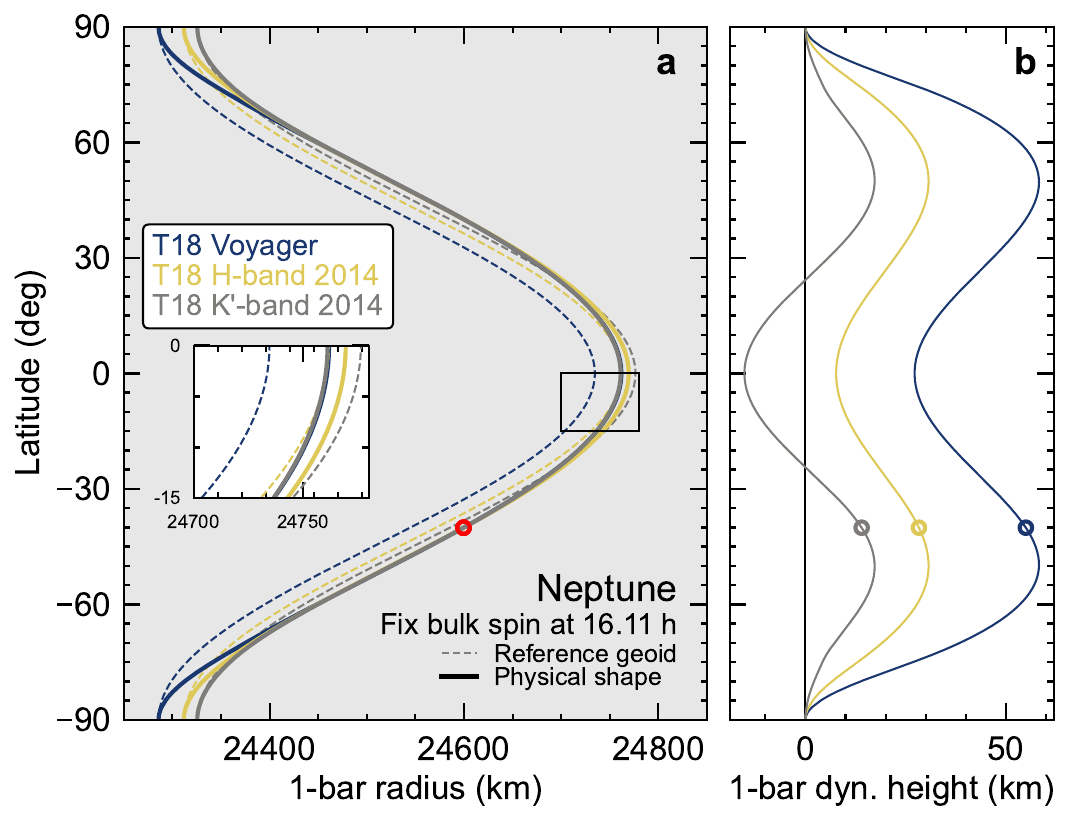}
        \caption{\label{fig.n_shape_vary_wind}
        As in Figure~\ref{fig.un_shapes}c-d, but holding Neptune's bulk rotation fixed at the 16.11 h magnetic field rotation period and considering a range of zonal wind profiles derived from Voyager and Keck data. These are the baseline profiles indicated in Figure~\ref{fig.wind_profiles}b.
        In light of the large variance in dynamical heights as a function of which wind profile is adopted (panel \textbf{b}), minimizing the mean dynamical heights would not yield a unique bulk rotation period until the wind profile is better known.
        }
    \end{center}
\end{figure}

Assuming that Uranus and Neptune's magnetic fields originate in fluid regions of their interiors, the considerable ($\sim2$~h) tension between these optimized periods and the magnetic rotation periods casts some doubt on the premise that minimizing the dynamical heights yields the correct interior rotation. If Jupiter and Saturn's winds tend toward the alignment of isobars and isopotentials, and Uranus and Neptune's do not, this underscores the potentially significant differences between the gas and ice giants vis-à-vis coupling of the atmosphere to the interior.

The shapes are also sensitive to the zonal wind profile, introducing nonuniqueness in the bulk rotation inferred from the condition $\langle h\rangle=0$. This effect is minor for Uranus but is more pronounced for Neptune, whose wind profile is more uncertain. Figure~\ref{fig.n_shape_vary_wind} shows that for fixed bulk rotation, the \cite{2018Icar..311..317T} wind fits yield dynamical heights that vary by up to 40~km depending on what fit is chosen. Optimizing the bulk rotation period such that $\langle h\rangle=0$ for each wind profile in turn produces periods ranging from 14.4 to 15.9 h. This exercise ignores the random spread in the statistical fits that \cite{2018Icar..311..317T} provide to each wind data set (the shaded regions in Figure~\ref{fig.wind_profiles}b give one example), which only exacerbate the non-uniqueness of any optimized Neptune spin period.

\section{Uncertainty of the shapes}\label{sec.uncertainties}
\begin{figure*}[ht!]
    \begin{center}
        \includegraphics[width=\textwidth]{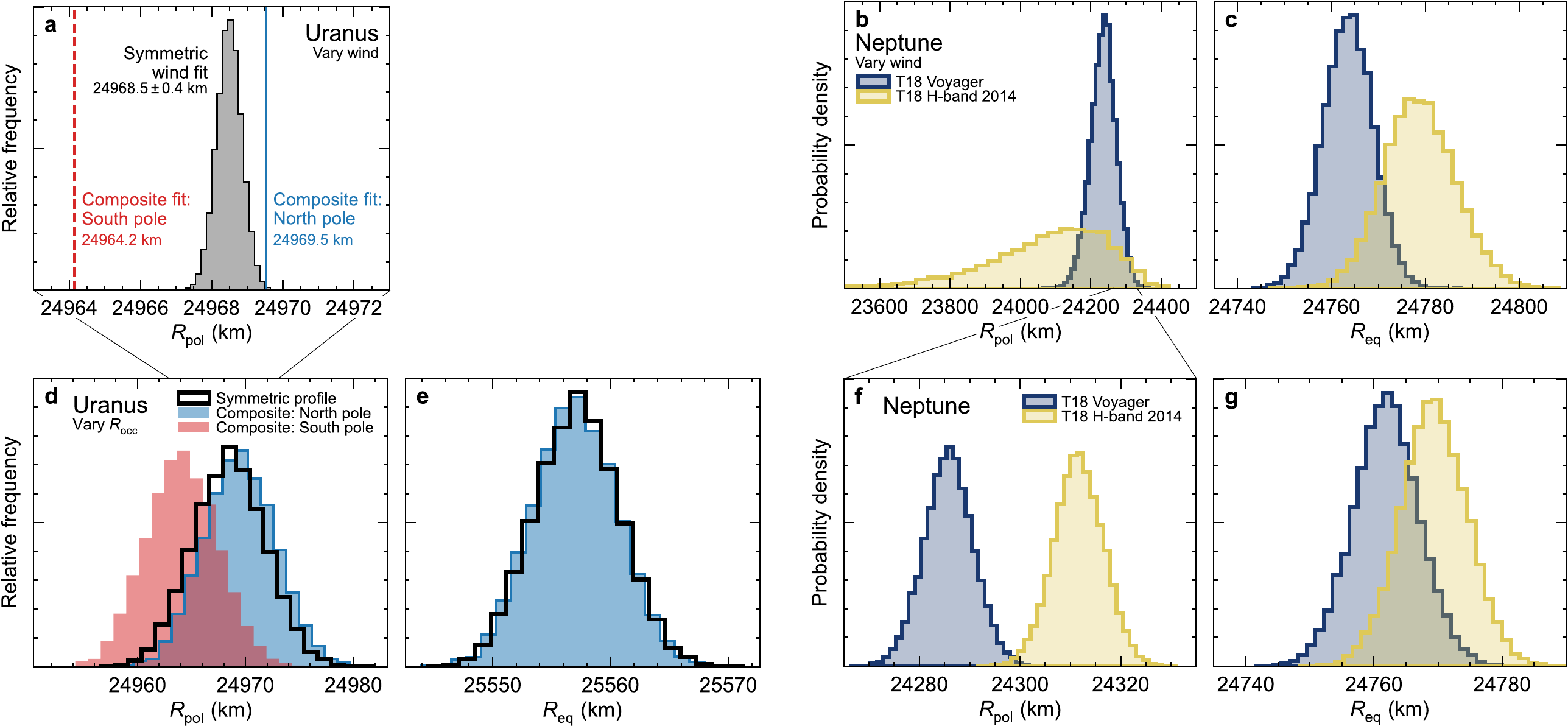}
        \caption{\label{fig.histograms}
        Distributions of the 1-bar polar and equatorial radii of Uranus and Neptune implied by uncertainties in the input wind profile (\textbf{a}-\textbf{c}) and the limited precision of radii inferred from radio occultations (\textbf{d}-\textbf{g}). Note the different scales in $\rpol$ between top and bottom rows.
        }
    \end{center}
\end{figure*}
\begin{figure}
    \begin{center}
        \includegraphics[width=\columnwidth]{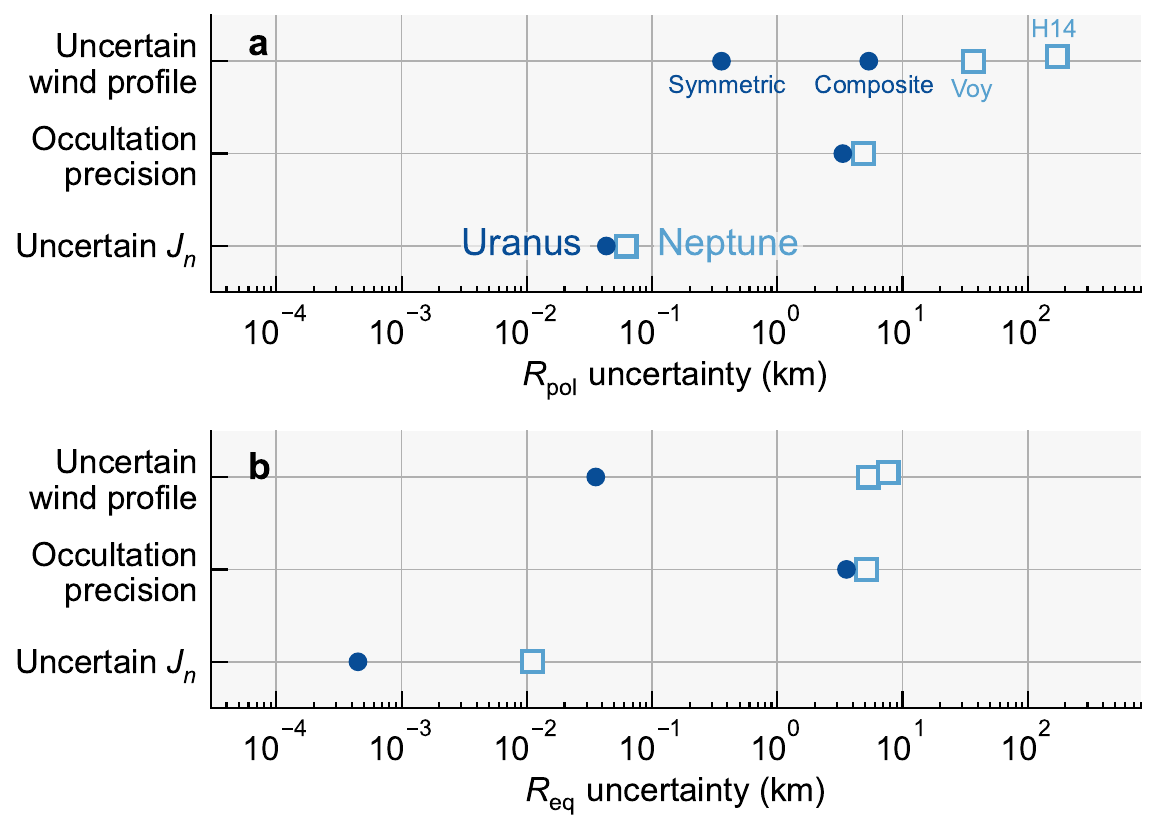}
        \caption{\label{fig.uncertainties}
        {A summary of the contributing factors to the overall uncertainty in Uranus and Neptune's polar radii (\textbf{a}) and equatorial radii (\textbf{b}). Radii correspond to the 1-bar isobaric surfaces of Uranus (dark circles) and Neptune (light squares) obtained by solving Equations~\ref{eq.gravity_radial}-\ref{eq.shape_ode}. The uncertainties in Uranus's polar and equatorial radii are both dominated by the $\sim5$~km radio occultation uncertainty. Uranus's possibly north-south asymmetric winds (labeled Composite for the corresponding wind profile in Figure~\ref{fig.wind_profiles}a-b) contribute to Uranus's $\rpol$ uncertainty at a similar level. For Neptune, the uncertain wind profile dominates the uncertainty in the polar and equatorial radii both.}
        }
    \end{center}
\end{figure}

As seen in Section~\ref{sec.shapes}, the shape solutions depend on three inputs derived from measurements:
{(1) the occultation radii, (2) the atmospheric rotation profile $\Omega(\phi)$, and (3) the zonal gravity moments $J_{n}$.} To quantify the present uncertainty in Uranus and Neptune's shapes, we vary one of these inputs at a time, arriving at the distributions of polar and equatorial radii presented in Figure~\ref{fig.histograms}.

\subsection{Occultation radii}\label{sec.uncertainties.occultations}
\cite{1987JGR....9214987L} and \cite{1992AJ....103..967L} do not explicitly quote uncertainties for the 1-bar radii derived from radio occultations. We take the radii to be normally distributed with standard deviation $\sigma_R=5$~km, matching the $5$-km navigation error cited by \cite{1981JGR....86.8721L} and similar to \cite{1987JGR....9214987L}'s quoted 4-km error on Uranus's 1-bar equatorial radius, just a few degrees in latitude from the Voyager 2 occultation. {We write a Gaussian log likelihood 
\begin{equation} 
    \ln P(\{R_i\}|\rpol)=-\sum_i\frac{1}{2\sigma_R^2}\bigg[R_i^{\rm model}-R_{i,\rm model}(\rpol)\bigg]^2,
\end{equation}
assign the single parameter $\rpol$ a prior probability $\ln P(\rpol)$ that is uniform over $[2, 3]\times10^4$~km, and sample from the posterior probability density $P(\rpol|\{R_i\})=P(\{R_i\}|\rpol)P(\rpol)$ using} \verb|emcee| {\citep{2013PSP..125..306F} \href{https://emcee.readthedocs.io/en/v3.1.6/}{v3.1.6}. The resulting $\rpol$ and $\req$ distributions are shown in Figure~\ref{fig.histograms}d-e for Uranus and Figure~\ref{fig.histograms}f-g for Neptune. Both $\rpol$ and $\req$ are Gaussian-distributed with standard deviation close to 5~km following from the precision of the occultation radii.
The Uranus wind profiles realized for these models show the largest scatter in wind speed near the maximum at $\pm55\deg$ ~m~s$^{-1}$, where the wind speed profiles show just 4~cm~s$^{-1}$ standard deviation. For the Neptune sample, wind speed variations are largest at the equator (40~cm~s$^{-1}$ standard deviation) and at $\pm75\deg$ (20~cm~s$^{-1}$) marking the onset of the high-latitude attenuation described in Section~\ref{sec.shapes}.
}

\subsection{Wind profiles}\label{sec.uncertainties.winds} 
{To estimate the distribution of shape solutions permitted by published wind profiles, we consider the statistical variation in the Uranus symmetric fit and the Voyager and H-band 2014 fits for Neptune (see Section~\ref{sec.wind_profiles}). 
Here again MCMC sampling is used to obtain posterior distributions, this time with a larger set of parameters comprising $\rpol$ plus a set of wind coefficients. These coefficients are 10 Legendre terms for Uranus (see \citealt{2018Icar..311..317T}'s Table 5, Fit 1) and 3 polynomial components for Neptune (\citealt{2018Icar..311..317T}'s Table 5, Fit 1). The coefficients are assumed independent and Gaussian-distributed. These experiments fix the occultation uncertainty $\sigma_R$ at $50$~m, an artificially small error serving to isolate the effect of the uncertain wind profile.
}

{For Uranus, the wind uncertainty yields $\rpol=24968.5\pm0.4$~km (Figure~\ref{fig.histograms}a), the standard deviation an order of magnitude smaller than what follows from the occultation uncertainty (Figure~\ref{fig.histograms}d). 
For comparison, Figure~\ref{fig.histograms}a also shows the north and south polar radii obtained when a single model assuming the asymmetric composite wind profile is fit to the Voyager occultations, a span of 5.3~km. This figure suggests that new occultations close to Uranus's pole may be able to discern between competing wind profiles, if precision on the order of 1~km is possible; see our Discussion below. The Uranus models show the largest model-to-model wind speed variation at the equator, where the standard deviation is 3~m~s$^{-1}$.
}

{For Neptune, the more uncertain winds are readily apparent in Figure~\ref{fig.histograms}b, the Voyager profile imparting a 37-km standard deviation in $\rpol$; for the H-band profile this rises to 170 km. The solutions' equatorial radii show smaller standard deviations of 5 and 8~km for the Voyager and H-band 2014 fits, respectively: Neptune's equatorial radius is better constrained owing to a greater density of feature-tracking observations close to the equator (per \citealt{2018Icar..311..317T} and the error band in Figure~\ref{fig.wind_profiles}c). 
This experiment produces wind speeds that vary widely, with the Voyager profiles producing standard deviation ranging from 1.5~m~s$^{-1}$ at the equator to 120~m~s$^{-1}$ at $\pm75\deg$. The H-band 2014 profiles yield standard deviation 30~m~s$^{-1}$ at the equator and 440~m~s$^{-1}$ at $\pm75\deg$.
}

\subsection{Gravity moments}\label{sec.uncertainties.jn}
{Finally, we quantify the influence of uncertain zonal gravity on the shape solutions. {Even the extreme occultation precision $\sigma_R=50$~m considered for the previous experiment is found to overwhelm the effect of varying $J_n$. We opt to forgo the MCMC formalism here and simply draw $J_n$ values and integrate Equation~\ref{eq.shape_ode} directly to the pole or equator, starting from whichever occultation measurement is closest. We draw $J_2$ and $J_4$ from multivariate normal distributions with mean and covariance taken from \cite{2024Icar..41115957F} for Uranus and \cite{2009AJ....137.4322J} for Neptune.} (We assume zero correlation between $J_2$ and $J_4$ for Neptune.) Although $J_6$ is not measured, we draw from a Gaussian with mean and standard deviation equal to $10^{-6}$ to very conservatively cover the distribution expected from interior models (e.g., \citealt{2022MNRAS.512.3124N}, \citealt{2024A&A...690A.105M}, \citealt{2025PSJ.....6...27L}, \citealt{2025mankovich_ugrav}). The $J_6$ uncertainty proves to dominate the $\rpol$ and $\req$ uncertainty for both planets, yielding an $\rpol$ standard deviation of 40 m (60 m) for Uranus (Neptune). The standard deviation in $\req$ is minor at 40 cm (10 m) for Uranus (Neptune). Variance in the wind speeds is negligible for these cases.
}

\subsection{Outlook}\label{sec.uncertainties.outlook}
Figure~\ref{fig.uncertainties} summarizes the uncertainties in the radii, plotting $1\sigma$ standard deviations for all distributions {obtained in Section~\ref{sec.uncertainties.occultations}-\ref{sec.uncertainties.jn}.}
The `composite' point in Figure~\ref{fig.uncertainties}a plots the total {5.3~km} pole-to-pole difference seen in Figure~\ref{fig.histograms}a.
For Uranus, {the statistical spread in the} symmetric wind fit yields a standard deviation of just $\sim400$~m, excluding the north polar radius predicted by the composite profile at nearly $3\sigma$ confidence and the south polar radius at $\geq21\sigma$ confidence. This significant separation indicates that the differences between different data sets, including possibly seasonally-dependent flow, are larger than the statistical spread in the velocities of features tracked within a single observation period.

The uncertainty in Uranus's polar radius is dominated by the $\sim5$~km radius precision achieved by the Voyager 2 radio occultation. This precision produces a $5$-km standard deviation in $\rpol$ for either choice of wind profile. (Note that uniform offsets to the occultation radii directly alter $\rpol$, which acts as an integration constant when integrating Equation~\ref{eq.shape_ode}.) The composite fit increases the north polar radius by approximately $1$~km, and predicts a uniform north-south polar radius contrast of $5$~km. From the significant overlap of all three distributions in Figure~\ref{fig.histograms}d, we conclude that radio occultations would need to achieve better than 5~km precision if new occultations near both poles are to discern between the symmetric and asymmetric wind models proposed for Uranus. Occultation radii precise to 1~km would suffice to test the pole-to-pole asymmetry implied by the composite profile. The wind profile has a negligible ($\sim2$~m) effect on Uranus's $\req$ because of the nearly equatorial geometry of the occultation. Uranus's uncertain $J_n$ affect $\rpol$ at the level of 40~m and $\req$ at less than 1~m.

For Neptune, the {dispersion in $\req$ is notably greater} than for Uranus, owing to Neptune's single Voyager 2 occulation falling at intermediate latitude $-40.1^\circ$. 
The uncertainty in Neptune's polar radius is dominated by the uncertain wind profile. Random $\rpol$ uncertainties range from 30~km to $150$~km, largest for \cite{2018Icar..311..317T}'s fit to Keck infrared data in H band. Hence, should a new spacecraft encounter Neptune, high-latitude radio occultations with Voyager-level precision could greatly improve our knowledge of Neptune's uncertain winds. 
Equatorial occultations would be of more marginal value in this regard, since the {5-8~km} variation induced by the uncertain winds is comparable to the 5~km working precision of the occultations.
As was found for Uranus, Neptune's uncertain $J_n$ are a minor influence, affecting $\rpol$ and $\req$ at the level of {60~m and 10~m}, respectively.

\section{Discussion}\label{sec.discussion}

Uranus's equatorial radius is constrained by the Voyager 2 occultation \citep{1987JGR....9214987L}, but its polar radius is not constrained directly. \cite{2010Icar..210..446H,2022MNRAS.512.3124N} have suggested that a near-polar radio occultation by an orbiter could probe Uranus's interior rotation {by effectively measuring the planetary oblateness, i.e., selecting among the reference geoids (dashed curves) of Figure~\ref{fig.u_shape_ignore_winds}a.} This would make it possible to test the magnetic rotation period as representing the planet's true interior rotation.
This test is compromised by Uranus's $\sim300$~m s$^{-1}$ zonal winds, however.
We argued in Section~\ref{sec.shapes} that the shape of the isobaric surface  encodes the \textit{total} rotation of that surface, regardless of any rigidly rotating reference geoid one might construct as one component of the total shape model. We conclude (Section~\ref{sec.basic_results}) that even a nearly polar radio occultation by an orbiter would not yield new information about Uranus's interior rotation, but could provide insight into the rotation of the atmospheric layers probed by the occultation {through the application of models like the ones developed here}.

{Occultation experiments close to the poles} would be valuable in their own right. They hold the potential to resolve open questions about the nature of Uranus's winds, particularly whether north/south asymmetries evident in the data are permanent or seasonal \citep{2024Icar..42016186S}. 
Provided that UOP arrives near Uranian equinox, both poles may accessible to occultations from a suitably inclined orbit. A profile based on Voyager 2 imaging of Uranus's southern latitudes and Keck/Gemini observations of northern and mid-latitudes \citep{2015Icar..250..294K,2015Icar..258..192S} predicts a 5~km difference between Uranus's northern and southern polar radii. This difference would not be discernible in polar radio occultations at the $\sim5$~km precision afforded by Voyager 2 radio science (Section~\ref{sec.uncertainties}), but there is reason to expect that UOP could achieve a precision of $1$~km or better.
For example, the $\approx5$~km precision on the radii for Voyager occultation of Jupiter is largely driven by the uncertainty in the spacecraft position and uncertainty in the shape itself, driven by the uncertainty in the gravity coefficients \citep{1981JGR....86.8721L}. We suspect a similar rationale motivates the claimed precision for the Uranus occultations. Spacecraft navigation has improved since the Voyager era and the gravity coefficients will improve significant with future missions (e.g., \citealt{2024PSJ.....5..116P});  hence, it is reasonable to expect this occultation error to drop significantly. If this is the case, errors in radii derived via radio occultation will be limited instead by thermal noise in the link Doppler residuals \citep{2025GeoRL..5213231C}. First-order error-propagation calculations suggest that a {properly-equipped Uranus orbiter---with a 3-4 meter high-gain antenna, 20-30 W X-band transmitter, and stable frequency reference---should achieve radio tracking accuracies comparable to previous missions. Appendix~\ref{app.ro_precision} presents an estimate of the precision on isobaric heights achievable by a Uranus orbiter. Barring an unusual geometry in the tour design, these retrieved occultation radii precisions should also be comparable to prior missions, e.g., $\sim0.4$~km achieved by Juno \citep{2026NatAs.tmp...29G}. 
}
This is partially a result of the fact that Uranus' colder atmosphere is more refractive than Jupiter's at the same pressure levels (refractive index $\propto 1/T$), offsetting somewhat the impact of poorer link signal-to-noise ratio from farther out in the solar system. Altogether, new occultations would be a powerful complement to continued measurements of Uranus's wind speeds from Earth, and ideally from imaging by UOP.


Neptune's zonal wind profile is much more uncertain, particularly toward the poles where trackable features are scant. The span of baseline profiles in Figure~\ref{fig.wind_profiles}d yields an $8$~km range in $\req$ and a $40$~km range in $\rpol$. Sampling from the full statistical fits to the same feature tracking data yield show that the dispersion in observed velocities lead to $\rpol$ and $\req$ uncertain at a level of at least $30$~km, and possibly in excess of 100~km. So long as this state of affairs lasts, any new measurements of Neptune's shape are more constraining of the planet's wind profile than of the interior rotation. Conversely, the continued effort to measure Neptune's atmospheric circulation from ground- or space-based observatories is the best means of improving our knowledge of Neptune's shape.

It is curious that Jupiter and Saturn's rotation periods yield small dynamical heights, but applying this condition to Uranus and Neptune yields rotation periods that may be unrealistically short (Section~\ref{sec.minimize_dynamical_heights}). If all or part of Uranus and Neptune's interiors rotate with 14$-$15~h periods, the lagging rotation of their magnetospheres at 16$-$17~h periods would present a major puzzle for theories of Uranus's internal structure and magnetic field generation. {It may be premature to rule out deep differential rotation in these planets, including for example shear between a convective shell undergoing dynamo action (see \citealt{2004Natur.428..151S,2006Icar..184..556S}) and a deeper frozen core \citep{2021PSJ.....2..222S} or stably stratified inner mantle \citep{2024PNAS..12103981M}.}
Rapid interior rotation would imply powerful retrograde equatorial jets in both planets (e.g., Figure~\ref{fig.wind_profiles}b). This situation is a departure from the paradigm of weakly retrograde equatorial jets implied by the ice giants' magnetospheric rotation periods, but notably one that emerges in hydrodynamical simulations as an outcome of Reynolds stresses associated with tilted convection columns \citep{2014MNRAS.438L..76G,2022ApJ...938...65C,2025SciA...11S8899D}.

If Uranus and Neptune's true rotation periods are {instead} consistent with their magnetospheric rotation {periods, Section~\ref{sec.minimize_dynamical_heights} suggests that their atmospheric mean dynamical heights are simply nonzero. Then} the question becomes, why are these planets' atmospheric dynamics different from Jupiter and Saturn's? Atmospheric modeling studies identify a number of contributing factors, including differences in the generation of Rossby waves and degree of angular momentum flux divergence close to the equator \citep{liu_2010, 2025SciA...11S8899D}, latent heating by condensation in the troposphere \citep{lian_2010}, the balance between buoyancy and Coriolis forces characterizing the deep convection \citep{2013Icar..225..156G,2014MNRAS.438L..76G}, or the depth to which the zonal winds penetrate \citep{guendelman_2025,2025SciA...11S8899D}. On this last point it interesting to note that Uranus's wind depth is already constrained somewhat by the planet's measured gravity moments; radio tracking of a Uranus orbiter could make a definitive measurement provided a suitable orbit \citep{2013Natur.497..344K,2020MNRAS.498..621S,2023AJ....165...27S,2024A&A...684A.191N,2025mankovich_ugrav}. 
In any case, consensus remains elusive when it comes to a general model that can consistently reproduce the observed zonal wind patterns of Jupiter, Saturn, Uranus, and Neptune simultaneously. The striking bifurcation phenomenon identified by \cite{2025SciA...11S8899D} may be an important aspect of the problem, but this and the aforementioned atmospheric models so far have emphasized the dichotomy between prograde and retrograde jets at the equator. It remains unclear how well these models reproduce Uranus or Neptune's prograde, mid- to upper-latitude jet streams which overwhelmingly determine the overall dynamical height of their atmospheres (see Figure~\ref{fig.un_shapes}).

\section{Conclusion}\label{sec.conclusion}
We applied standard geodetic calculations to study the influence that Uranus and Neptune's interior rotation, gravity fields, and zonal winds have on these planets' shapes. In an idealized case where the gravity field and total atmospheric rotation profile are perfectly known, a radius constraint from a single radio occultation is sufficient to uniquely determine the shape of the 1-bar isobaric surface. Changing the interior rotation of the planet has no direct effect on this solution, provided that the atmospheric rotation is preserved. 

For Uranus and Neptune, whose radii are constrained by Voyager~2 radio occultations and whose atmospheric rotation profiles are not fully understood, a family of solutions is left open for each planet's 1-bar isobaric shape. For Uranus, the $\sim5$~km precision of the available occultation radii make it difficult to discern between competing models for Uranus's atmospheric flow. The presence of a persistent north-south asymmetry in the flow would be testable by UOP if radio occultations can be carried out close to the poles (preferably both poles) and the 1-bar radius can be measured to within $1$~km precision. For Neptune, the uncertainty in the shape is overwhelmed by the uncertain atmospheric rotation profile. Occultations by a future Neptune orbiter would be a valuable means of constraining Neptune's atmospheric rotation to complement measurements made remotely using feature tracking.

These findings motivate future studies to improve the precision of isobaric radii derived from spacecraft radio links, and new observations to better quantify the zonal wind profiles of Uranus and Neptune (from the ground, and additionally from UOP in the case of Uranus). For Uranus, testing the hypothesis that the interior rotates with the well-constrained magnetic rotation \citep{2025NatAs...9..658L} might be best served by alternative strategies such as ring or Doppler imaging seismology \citep{2025PSJ.....6...70M} or gravitational seismology \citep{2020RSPTA.37890475F,friedson26,2025parisi_uranus_grav_seismology}.

\section*{Acknowledgments}
    We thank Eli Galanti and an anonymous reviewer for their helpful insights.
    A portion of C.M.'s research was supported by an appointment to the NASA Postdoctoral Program at the Jet Propulsion Laboratory, administered by Oak Ridge Associated Universities under contract with NASA.
    JPL internal funding is acknowledged.
    The research described in this paper was carried out at the Jet Propulsion Laboratory, California Institute of Technology, under a contract with the National Aeronautics and Space Administration (80NM0018D0004). 
    \copyright 2026 California Institute of Technology. Government sponsorship acknowledged.


\appendix
\allowdisplaybreaks
\section{Estimating the Precision of Occultation-Derived Pressure Heights at Uranus}\label{app.ro_precision}
We provide a first-order estimate for the precision of occultation-derived pressure heights at Uranus via propagation of errors assuming a spherically-symmetric geometry with the occultation plane defined by the 
Earth-Uranus vector and the velocity vector of a spacecraft in a circular orbit (see \href{https://iopscience.iop.org/article/10.3847/PSJ/accae3}{Akins et al. 2023} for a more complete description). This geometry establishes simple relationships between the observed Doppler shift $f$, the ray bending angle $\delta$, and the ray impact parameter $a$, facilitating error propagation \citep{1973P&SS...21.1521E}

\begin{equation} \label{eq:fres}
f = (v_t / \lambda_0) \sin{\delta},
\end{equation}

\begin{equation} 
a = R_U \cos{(\gamma - \delta)}.
\end{equation}

Here, $v_t$ is the spacecraft velocity (in the principal axis of the occultation plane), $\lambda_0$ is the link wavelength, $\gamma$ is the complement angle of the spacecraft elevation with respect to the Earth-Uranus vector, and $R_U$ is the distance of the spacecraft from the center of Uranus. The relationship between impact parameter and atmospheric refractive index $n$ is given by the usual inverse Abel transform

\begin{equation} \label{eq:refractderiv}
\ln{n(a)} = \frac{1}{\pi} \int_{a}^\infty \frac{\delta da'}{\sqrt{a'^2-a^2}}.
\end{equation}

The refractivity is linearly related to temperature and pressure as $(n-1) \times 10^6 = C(p/p_0)(T_0 / T)$, where $C=$ 135.77, 34.51, and 440 for H$_2$, He, and CH$_4$ at $p_0=1$ atm and $T_0=273.15$ K. While this relationship relates occultation refractivity uniquely to atmospheric density, temperature and pressure profiles as a function of altitude can be separated by assigning top-of-atmosphere values and integrating the hydrostatic balance equation downward. 

Further simplifications in our analysis are motivated by the recent Juno results \citep{2025GeoRL..5213231C}.
Their analysis demonstrates that, towards the denser 1-bar pressure level, thermal noise on the measured Doppler shift is the dominant contribution to errors in retrieved refractivity. Uncertainties in wind profile, relevant to their ray-tracing procedure, also begin to become relevant at this altitude; we neglect these here and simply note that these errors should be at most approximately equal to the thermal noise errors at 1 bar. 

We first compute atmospheric refractivity profiles using an equilibrium cloud condensation model \citep{1973Icar...20..465W}.
We then compute Doppler uncertainty using the standard radio link thermal noise relationship, which is a function of signal bandwidth $B$, carrier-to-noise ratio $C/N_0$, and integration time $\tau=1$ s:

\begin{equation}
\sigma^2_{f} = \frac{2B [N_0/C]}{(2\pi \tau)^2}.
\label{eq:sigmaf}
\end{equation}

\begin{figure}[htbp]
\begin{center}
\includegraphics[width=0.5\linewidth]{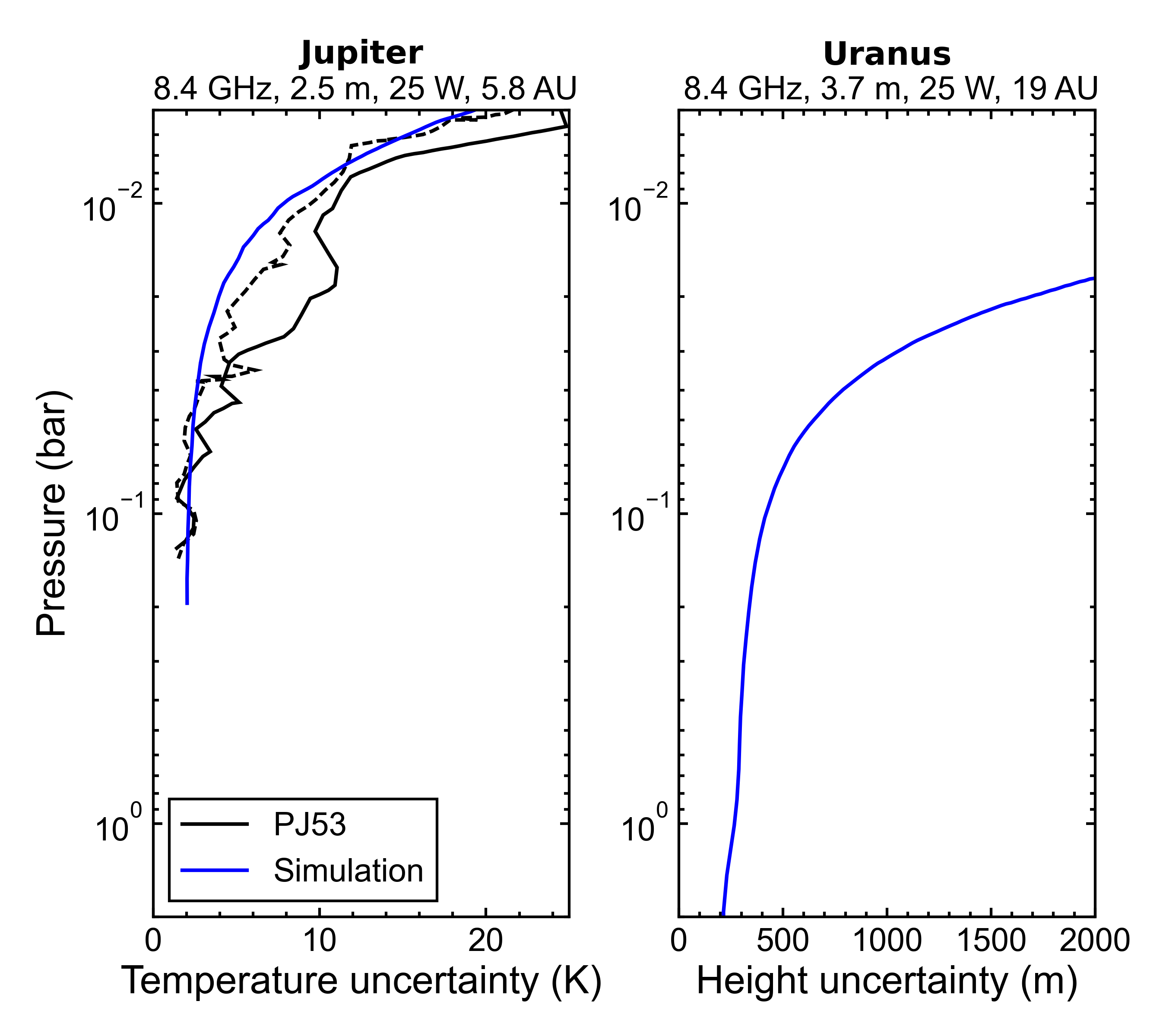} 
\caption{Simulation results, with link frequency, spacecraft high-gain antenna diameter, transmitter power, and distance from Earth shown in the titles. (Left) Simulated uncertainty in temperature for radio occultations of Jupiter compared with the Juno PJ53 results from Caruso et al. (2025). (Right) Calculation of pressure-height uncertainty for a nominal Uranus orbiter. }
\label{fig.uranus_occultation_precision}
\end{center}
\end{figure}

We propagate errors through the relations above, using the finite discretization approach of \cite{1979Icar...39..192L}
to propagate errors through the inverse Abel transform. To counteract the simplifications in our approach, we further scale the Doppler uncertainty term such that, using a Jupiter model atmosphere and Juno radio science link parameters, the propagation of errors for temperature matches the \cite{2025GeoRL..5213231C} results for PJ53. Applying this scaling factor then to simulations for Uranus, and computing instead the uncertainty in assigned geopotential height, we find that $<$ 1 km precision should be achievable for a Voyager-class orbiter (see Figure~\ref{fig.uranus_occultation_precision}), consistent with the 400-m errors achieved at Jupiter by \cite{2026NatAs.tmp...29G}.
 
\section{Shape Model Validation at Jupiter and Saturn}\label{app.jupiter_and_saturn}

\begin{figure*}
    \begin{center}
        \includegraphics[width=\textwidth]{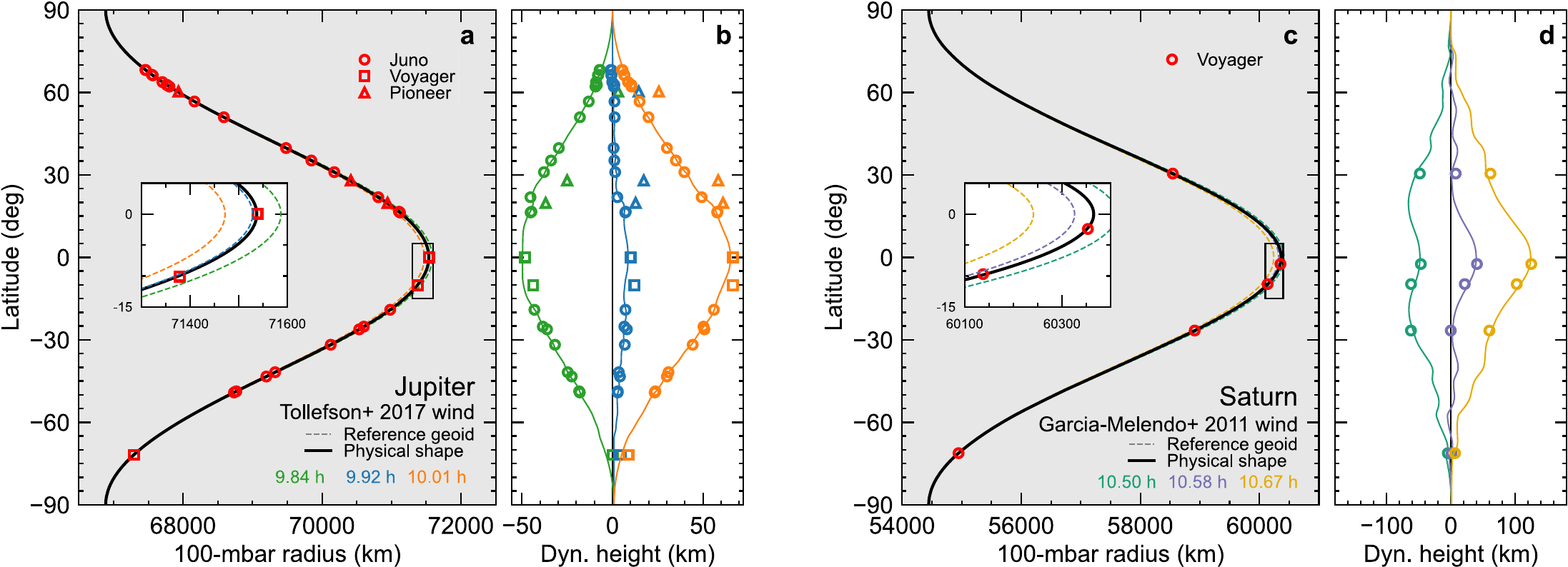}
        \caption{\label{fig.js_shapes}
        As in Figure~\ref{fig.un_shapes}, but for the 100-mbar level in Jupiter (\textbf{a}-\textbf{b}) and Saturn (\textbf{c}-\textbf{d}). 
        }
    \end{center}
\end{figure*}

{We apply the method of Section~\ref{sec.shapes} to Jupiter and Saturn, arriving at the shape solutions shown in Figure~\ref{fig.js_shapes}.}
We considered Jupiter and Saturn zonal wind profiles from \cite{2017Icar..296..163T} and \cite{2011Icar..215...62G} respectively. Wind velocities are set to zero poleward of 80 deg latitude for Saturn. 100-mbar occultation radii are from \cite{1981JGR....86.8721L}, \cite{2026NatAs.tmp...29G} and \cite{1985AJ.....90.1136L}. 
{As for Uranus and Neptune, solutions are obtained by finding the $\rpol$ value that minimizes the RMS radius residual with respect to occultation data. For Jupiter, Juno occultation results are taken from \cite{2026NatAs.tmp...29G} and Pioneer and Voyager results are taken from \cite{2009P&SS...57.1467H}'s estimates based on figures in \cite{1985AJ.....90.1136L} and \cite{1981JGR....86.8721L}. For Jupiter, the Pioneer and and Voyager points show larger offsets with respect to our favored shape solution, and we opt to retain only the Juno points in our fit. Good agreement is observed, with an RMS error of 0.7~km for Jupiter (4.2~km including the Voyager and Pioneer points), and 4.2~km for Saturn. We obtain $\rpol=66,884.9$~km and $\req=71,537.2$~km for Jupiter's 100-mbar surface, in excellent agreement with \cite{2026NatAs.tmp...29G}. For Saturn's 100-mbar surface, we find $\rpol=54,445.7$~km and $\req=60,365.4$~km, in agreement with \cite{1985AJ.....90.1136L}'s original shape solution to within their reported uncertainties. This polar radius exceeds \cite{2023GeoRL..5002321G}'s favored value for Saturn by $9$~km, in part a consequence of their gravity-optimized wind profile as opposed to the unaltered \cite{2011Icar..215...62G} profile considered here for simplicity.}

Reference geoids and dynamical heights are shown for a $\pm5$~min range of rigid rotation periods, centered at Jupiter's System III period 9h 55m 29.71s (\citealt{2018CeMDA.130...22A}; see also \citealt{2011JGRA..116.5217H}) and at 10h 35m for Saturn. In Figure~\ref{fig.js_shapes}d we recover the known result that Saturn's dynamical heights are minimized for rotation periods between 10h~30m and 10h~35m \citep{2007Sci...317.1384A,2009P&SS...57.1467H}, in good agreement with estimates of Saturn's interior rotation based on atmospheric {Rossby waves} \citep{2009Natur.460..608R} and ring seismology \citep{2023PSJ.....4...59M}. Similarly, Jupiter's precisely known radiometric rotation period does yield mean atmospheric dynamical height close to zero (see Section~\ref{sec.minimize_dynamical_heights} and \citealt{2009P&SS...57.1467H,2023GeoRL..5002321G,2026NatAs.tmp...29G}); $\langle h\rangle=3.1$~km for the central model in Figure~\ref{fig.js_shapes}a-b. {Like the Uranus and Neptune models described in Section~\ref{sec.shapes}, these Jupiter and Saturn isobaric surfaces and reference geoids are available in plaintext format as part of the Github and Zenodo repositories referenced in the main text.}

\end{document}